\documentclass[12pt,preprint]{aastex}

\slugcomment{}

\shorttitle{A Possible New Distance Indicator}
\shortauthors{Sato et al.}

\begin{document}


\title{A Possible New Distance Indicator\\-Correlation between the duration and the X-ray luminosity of the shallow decay
 phase of Gamma Ray Bursts-}


\author{
   R. Sato\altaffilmark{1},
   K. Ioka\altaffilmark{2},
   K. Toma\altaffilmark{2},
   T. Nakamura\altaffilmark{2},
   J. Kataoka\altaffilmark{3},
   N. Kawai\altaffilmark{3}, and 
   T. Takahashi\altaffilmark{1}}


\altaffiltext{1}{Institute of Space and Astronautical Science/JAXA, Sagamihara, Kanagawa 229-8510, Japan}
 \email{rsato@astro.isas.jaxa.jp}
 \altaffiltext{2}{Department of Physics, Kyoto University, Kitashirakawa, Sakyo-ku, Kyoto 606-8502, Japan}
 \altaffiltext{3}{Department of Physics, Tokyo Institute of Technology, Meguro-ku, Tokyo 152-8551, Japan}


\begin{abstract}
We investigated the characteristics of the shallow decay phase in the early X-ray afterglows of GRBs 
observed by $Swift$ X-Ray Telescope (XRT) during the period of January 2005 to December 2006. 
 We found that the intrinsic  break time at the shallow-to-normal decay transition in the X-ray light curve 
$T^0_{\rm brk}$ is moderately well correlated with the isotropic X-ray luminosity 
in the end of the shallow decay phase ($L_{X, \rm end}$) as 
$T^0_{\rm brk} = (9.39\pm0.64) \times 10^3 {\rm s} ({L_{X,\rm end}}/{10^{47} {\rm ergs\cdot s^{-1}}})^{-0.71\pm0.03}$, 
while $T^0_{\rm brk}$ is weakly  correlated with the isotropic gamma-ray energy of the prompt emission $E_{\gamma, \rm iso}$.
Using  $T^0_{\rm brk}-L_{X, \rm end}$ relation we have determined the pseudo redshifts of 33 GRBs. 
We compared the pseudo redshifts of 11 GRBs with measured redshifts and found the rms error to be 0.17 in $\log z$. 
From this pseudo redshift, we estimate that $\sim 15\%$ of the $Swift$ GRBs have $z > 5$.  
 The advantages of this distance indicator is that
(1) it requires only  X-ray afterglow data while other methods such as Amati and Yonetoku correlations 
require the peak energy ($E_p$) of the prompt emission, 
(2) the redshift is uniquely determined without redshift 
degeneracies unlike the Amati correlation,
and (3) the redshift is estimated in advance of deep follow-ups
so that possible high redshift GRBs might be selected for detailed observations.
\end{abstract}





\section{Introduction}

The optical afterglow light curves of most GRBs show the smooth power-law decays with time ($\propto t^{-\alpha}$ 
with a typical index value $\alpha \sim 1$). 
This is consistent with the prediction of the simplest
model of a spherical blast wave propagating into a uniform medium,
where the spectrum consists of several power-law segments in which
$F_{\nu}\propto\nu^{-\beta}t^{-\alpha}$ (Sari, Piran \& Narayan 1998).
 
However, after the advent of the $HETE$-2 satellite,  
prompt localization of GRBs sent to ground-based telescopes made it possible  
to observe the early afterglows which revealed that
some GRBs show deviations from a smooth power-law light curve.
For example, GRB 021004 showed a highly variable light curve, with several bumps 
and wiggles from a simple power-law (e.g., Fox et al. 2002; Mirabal et al. 2002).
GRB 030329 also showed such a short timescale variability (Uemura et al. 2003; Sato et al. 2003).
Furthermore, the early slow decay was observed in GRB 021004 and was interpreted by Fox et al. (2003)
as arising from delayed shocks and continuous energy ejection from the central engine 
(Rees \& Meszaros 1998; Kumar \& Piran 2000; Sari \& Meszaros 2000).

The early X-ray afterglow was found to be even more complex by the $Swift$ observations. 
They generally consist of three distinct power-law segments 
(Nousek et al. 2006; Zhang et al. 2006):
(1) an initial steep decay with $\alpha_1 \sim 3$, 
(2) a shallow decay with $\alpha_2 \sim 0.5$ and finally 
(3) a normal decay $\alpha_3 \sim 1$, where $\alpha_1$, $\alpha_2$ and $\alpha_3$ are power-low 
indices of temporal variability ($\alpha$ of $t^{-\alpha}$). 
The initial steep decay is most likely the transition from the prompt emission 
to the afterglow (Nousek et al. 2006; Zhang et al. 2006). 
Such a steep decay is due to the ``curvature effect'' of the high-latitude emission 
expected for the emission ceasing abruptly (e.g., Kumar \& Panaitescu 2000; Dermer 2004; Dyks et al. 2005; 
Panaitescu et al. 2006; Yamazaki et al. 2006). 
In the normal decay phase, the data for many bursts are consistent with 
an ISM medium rather than a wind medium. 
However, for the shallow decay phase, its physical background has not been understood 
and it is the most mysterious feature in the early X-ray afterglows.

As for shallow decay phase,
Willingale et al. (2007) have studied the end-time of the shallow decay 
of the X-ray light curves of long GRBs, $T_{\rm a}$. 
They suggested a possibility that $T_{\rm a}$ depends on the total energy of the outflow.
Nava et al. (2007) have investigated the correlation between 
the end-time in the GRB frame $T_{\rm a}^{0} \equiv T_{\rm a}/(1 + z)$ 
and the isotropic gamma-ray energy of the prompt emission $E_{\gamma, \rm iso}$.
They found that for the bursts in their sample, $T_{\rm a}^0$ is weakly correlated with 
$E_{\gamma,\rm iso}$. However, this correlation disappears
when considering all bursts of known redshift and $T_{\rm a}$.
Liang et al. (2007) have also showed that there is no significant correlation 
between the break time between shallow and normal decay segments and the $E_{\gamma, \rm iso}$.

In this paper we investigated the characteristics of the shallow decay phase in the early X-ray afterglows of GRBs 
observed by $Swift$ X-Ray Telescope (XRT) during the period of January 2005 to December 2006. 
We found that the intrinsic break time at the shallow-to-normal decay transition in the X-ray light curve 
$T^0_{\rm brk}$ is correlated with the isotropic X-ray luminosity of the shallow decay phase, $L_{\rm X}$. 
We tried to apply this relation to determine the redshifts of 33 GRBs and found that the distribution 
of the pseudo redshifts is similar to that of spectroscopically measured redshifts with more high pseudo redshift GRBs. 
Furthermore, we independently examine if there are any correlations among parameters 
of the prompt emission and the shallow decay phase for GRBs. 
We found that $T^0_{\rm brk}$ is weakly correlated with the isotropic gamma-ray energy of the prompt emission 
$E_{\gamma, \rm iso}$.

Throughout this paper, we adopt the cosmological parameters
$H_0 = 71$ km s$^{-1}$ Mpc$^{-1}$, $\Omega_{\rm M} = 0.27$ and $\Omega_{\Lambda} = 0.73$.

\section{Data selection and analysis}

In this section, we present the data samples and the calculation method of parameters as 
$L_{\rm X}$, $E_{\rm \gamma, iso}$  and $T_{\rm brk}$. 
To examine the $L_{X}$ and $T^0_{\rm brk}$ relation, the measured redshift $z$ and XRT data 
at the start and the end time ($T_{\rm brk}$) of the shallow decay phase are needed: 
The isotropic X-ray luminosity $L_{\rm X}$ is calculated from the X-ray spectrum and 
the break time at the shallow-to-normal decay transition $T_{\rm brk}=(1+z)T^0_{\rm brk}$ 
is obtained from the X-ray light curve. 
Here we define the shallow decay phase when the light curve decay index is flatter 
than the canonical value of $\alpha \sim 1$.
We found that 11 GRBs have well defined measured values of these parameters between January 2005 and December 2006. 

On the other hand, in order to examine the $E_{\gamma,\rm iso} - T^0_{\rm brk}$ relation, 
the redshift $z$, the photon energy at the peak of the $\nu F_{\nu}$ spectrum, $E_{\rm p}$ for calculating 
$E_{\gamma, \rm iso}$ and $T_{\rm brk}$ are needed. $E_{\rm p}$ is obtained from the GRB spectrum. 
A GRB spectrum is typically described by a Band function (Band et al. 1993), 
which is a smoothly-joint broken power-law characterized by two photon indices and $E_{\rm p}$. 
We found that seven GRBs had measured value of these parameters between January 2005 and December 2006. 
$E_{\rm p}$ for these GRBs has been firmly identified from the $Konus$, $HETE-2$ observations reported in Amati et al. (2006b).
Generally it is difficult to determine $E_{\rm p}$ with $Swift$-BAT data alone due to its narrow energy range of $15-150$ keV.
For the GRBs whose $E_{\rm p}$ cannot be determined, we have developed a method to estimate $E_{\rm p}$ using 
the $E_{\rm p} - L_{\gamma, \rm iso}$ correlation (Yonetoku et al. 2004; Ghirlanda et al. 2005) 
and additionally obtained $E_{\rm p}$ for 11 GRBs including GRB 050824 which has only the lower limit for $T_{\rm brk}$. 

In this section we systematically analyze the XRT and the Burst Alert Telescope (BAT) data 
and show the details of calculation methods to obtain the parameters of $L_{\rm X}$, $E_{\rm \gamma, iso}$ and $T_{\rm brk}$. 

\subsection{XRT analysis}

\subsubsection{Reduction}

The elapsed time of the events such as the break time is measured from the BAT trigger time in this paper.
The XRT data presented here were obtained using the Window Timing (WT)
and/or the Photon Counting (PC) modes (event grades 0$-$2 and 0$-$12 respectively).
The XRT data were processed using $xrtpipline$ into filtered event lists.
Data were also filtered to eliminate time periods when the CCD temperature
was warmer than $-50^{\circ}$C. 
These filtered data were then used to extract light curves and spectrum in the 0.5$-$10 keV energy range.
For the PC data, the light curves and the spectrum are generally extracted 
from a circular region with a radius of 47'' (the region size depends on Point Spread Function (PSF) 
of the X-ray afterglows.).
The backgrounds are selected from an annulus region with radii of 94'' and 188'', 
excluding the X-ray source region near the GRB position. 
For the WT data, the light curves and the spectrum are extracted from a rectangular region of 94'' by 47''. 
The background is selected from a rectangle of the same size as for the source region 
that is typically 47'' away from the GRB position. 
In all cases, we used XSELECT to extract source and background and XSPEC version 11.3.2 to fit the spectra. 
 
The data obtained in the PC mode sometimes suffered from pile-up 
when the observed sources were brighter than 0.5 cts/s.
These data were corrected for pile-up by adopting the method described in Vaughan et al. (2006).
We used an annular extraction region, with a 9.4'' inner radius and a 47'' outer radius. 
In order to determine the correction factor for the annular aperture,
we modeled the PSF using XIMAGE.
The effective area is corrected using the calibration data and $xrtmkarf$.

\subsubsection{Estimation of $T_{\rm brk}$}

We consider 21 GRBs with known redshift and identified the shallow decay phase in the X-ray light curves. 
In order to determine $T_{\rm brk}$, we fitted the X-ray light curve obtained in Section 2.1.1 to a broken power-law model:
\begin{equation}
N(t) = \left\{
\begin{array}{@{\,}ll}
N_0 \times (t-t_0)^{-\alpha_1} \hspace*{5.15cm} ( t \leq T_{\rm brk} ),\\
N_0 \times (T_{\rm brk}-t_0)^{-\alpha_1} \times
\biggl(\frac{t-t_0}{T_{\rm brk}-t_0}\biggr)^{-\alpha_2} \hspace*{1.8cm} ( t > T_{\rm brk} ),
\end{array}
\right .\
\end{equation}
where $N_0$ is the normalization in units of counts s$^{-1}$, 
$\alpha_1$ and $\alpha_2$ are the temporal power-law indices before and after
the break time $T_{\rm brk}$. 
The temporal decay slopes and the break times of the shallow-to-normal transition 
are summarized in Table \ref{table:LCfit}. 
These parameters were determined using the BAT trigger time as $t_0$.

\subsubsection{Estimation of $L_{\rm X}$}

In order to determine $L_{\rm X}$, we fitted the XRT spectra obtained in Section 2.1.1 with the single power-law 
($N(E) \propto E^{-\Gamma}$, where $\Gamma$ is the differential photon index). 
We here adopt a  k-correction to $L_{\rm X}$. 
Let us write $N(E)dE=AE^{-\Gamma}dE$, then $L_X$ integrated from $E^0_d$ to $E^0_u$ in the GRB frame is given by 
\begin{equation}
 L_X=4\pi d_L(z)^2(1+z)^{\Gamma-2} A \frac{(E^0_u)^{2-\Gamma} - (E^0_d)^{2-\Gamma}}{(2-\Gamma)},
\end{equation}
where $d_{\rm L}$ is the luminosity distance, $\Gamma$ is the power-low index, $E^0_d$ and  $E^0_u$ are measured in the GRB frame. 
To determine $E^0_d$ and  $E^0_u$ we consider GRB at the Swift average of $z \sim 3$ 
observed in $0.5-10$ keV band so that  $E^0_d=2$ keV and  $E^0_u=40$ keV. 
We tested for the X-ray luminosity at initial ($L_{X, \rm ini}$), median ($L_{X, \rm med}$) 
and end ($L_{X, \rm end}$) of the shallow decay light curve. 
The results are summarized in Table \ref{table:Lx}. 
Some events did not have observation at beginning of the shallow decay phase 
for convenience of Swift observation. For these events, we show only the parameters 
at the end of the shallow decay light curve. 

\subsection{BAT analysis}

\subsubsection{Reduction} 

We analyze the BAT data of the $Swift$ GRBs observed during the period January 2005 to December 2006. 
The BAT data for the GRB samples were processed using standard $Swift$-BAT analysis software 
as described in the BAT Ground Analysis Software Manual ($http://heasarc.gsfc.nasa.gov/docs/swift/analysis/$).
Each BAT event was mask-tagged using $batmaskwtevt$ task with the best fit source position.
All of the BAT spectra have been background subtracted with this method.

We extracted the spectra in the energy range $15-150$ keV 
over the period corresponding to T$_{90}$ excluded during slew. 
If the satellite started to slew during T$_{90}$, we extracted the
spectrum before the slew start times.
If the satellite did not slew during T$_{90}$, we extracted the
spectrum over the whole T$_{90}$.
All spectra were fitted with XSPEC version 11.3.2.
The detector responses are generated from $batmaskevt$.

\subsubsection{Estimation of $E_{\gamma, \rm iso}$}

In this section, we consider 11 $Swift$ GRBs for which spectroscopic redshifts 
and $T_{\rm brk}$ are both available.
A GRB spectrum is typically described by a Band function (Band et al. 1993). 
The typical values of two photon indices are $\Gamma_1 \sim 1.0$ and $\Gamma_2 \sim 2.2$, respectively. 
However, the spectral peak energy values of GRB is typically $E_{\rm p}\sim250$ keV, i.e., 
above the BAT energy band (15$-$150 keV). Actually most spectra observed by the BAT 
are well fitted by a single power-law function. 
Generally, with the BAT observations alone, we cannot determine $E_{\rm p}$ and the high energy photon index $\Gamma_2$.

We estimate $E_{\rm p}$ by using the $E_{\rm p} - L_{\gamma, \rm iso}$ relation 
(Yonetoku et al. 2004; Ghirlanda et al. 2005) where $L_{\gamma, \rm iso}$ is 
the gamma-ray isotropic luminosity 
\footnote{The $E_{\rm p} - E_{\gamma,\rm iso}$ Amati relation (Amati et al. 2002; Friedman \& Bloom 2005; Amati 2006) 
is a similar luminosity relation. However, this does not produce a unique or necessarily 
well-determined $E_{\rm p}$ value 
because there is an intrinsic redshift degeneracy in the Amati relation (Li 2006; Schaefer \& Collazzi 2007). }.
Since the photon index obtained from the BAT is distributed $< 2$ in most case, 
the peak energy $E_{\rm p}$ of bursts is expected to be above the BAT energy band.
Therefore, we assume a broken power-law shape for the spectra time-averaged over the GRB duration: 
\begin{equation}
N(E) = \left\{
\begin{array}{@{\,}ll}
A \times E^{-\Gamma_1} \hspace*{3.7cm} {\rm for} \hspace*{0.3cm}  E\leq E_{\rm p},\\
A \times E_{\rm p}^{-(\Gamma_1 - \Gamma_2)} E^{-\Gamma_2} \hspace*{2cm} {\rm for} \hspace*{0.3cm} E> E_{\rm p}.\\
\end{array}
\right .\
\end{equation}
where $A$ is a normalization, $\Gamma_1$ and $\Gamma_2$ are the low and high energy photon indices, 
respectively, and $E_{\rm p}$ is the peak energy.
$A$ and $\Gamma_1$ can be determined from BAT data, and $\Gamma_2$ is assumed to have the typical value of 2.2.
The time-averaged flux in a given bandpass ($E_1$, $E_2$), where $E_1$ and $E_2$ are the minimum and the maximum energy, 
respectively, as a function of $E_{\rm p}$ can be given as:
\begin{equation}
F_{E_1 - E_2}(E_{\rm p}) = \int_{E_1}^{E_2} EN(E)dE.
\end{equation}
Therefore, the isotropic luminosity $L_{\gamma, \rm iso}$ is calculated as
\begin{eqnarray}
L_{\gamma, \rm iso} &=& 4\pi d_{\rm L}^2 \times F_{E_1 - E_2}(E_{\rm p}), \\
            &\sim& 4\pi d_{\rm L}^2 \times A \biggl( \frac{1}{2-\Gamma_1} - \frac{1}{2-\Gamma_2} \biggr) E_{\rm p}^{2-\Gamma_1}.
\end{eqnarray}
where $d_{\rm L}$ is the luminosity distance.

Meanwhile, the $E_{\rm p} - L_{\gamma, \rm iso}$ correlation was proposed by Yonetoku et al (2004). 
Ghirlanda et al. (2005) re-examined this correlation with an enlarged sample 
and showed the correlations with smaller scatter:  
\begin{equation}
\frac{E_{\rm p}}{\rm 100 \hspace*{0.2cm} keV} = (4.88\pm0.06)\times\biggl(\frac{L_{\gamma, \rm iso}}{1.9\times10^{52}}\biggr)^{0.48\pm0.01}.
\end{equation}
 
\begin{figure}
\begin{center}
\includegraphics[scale=.50]{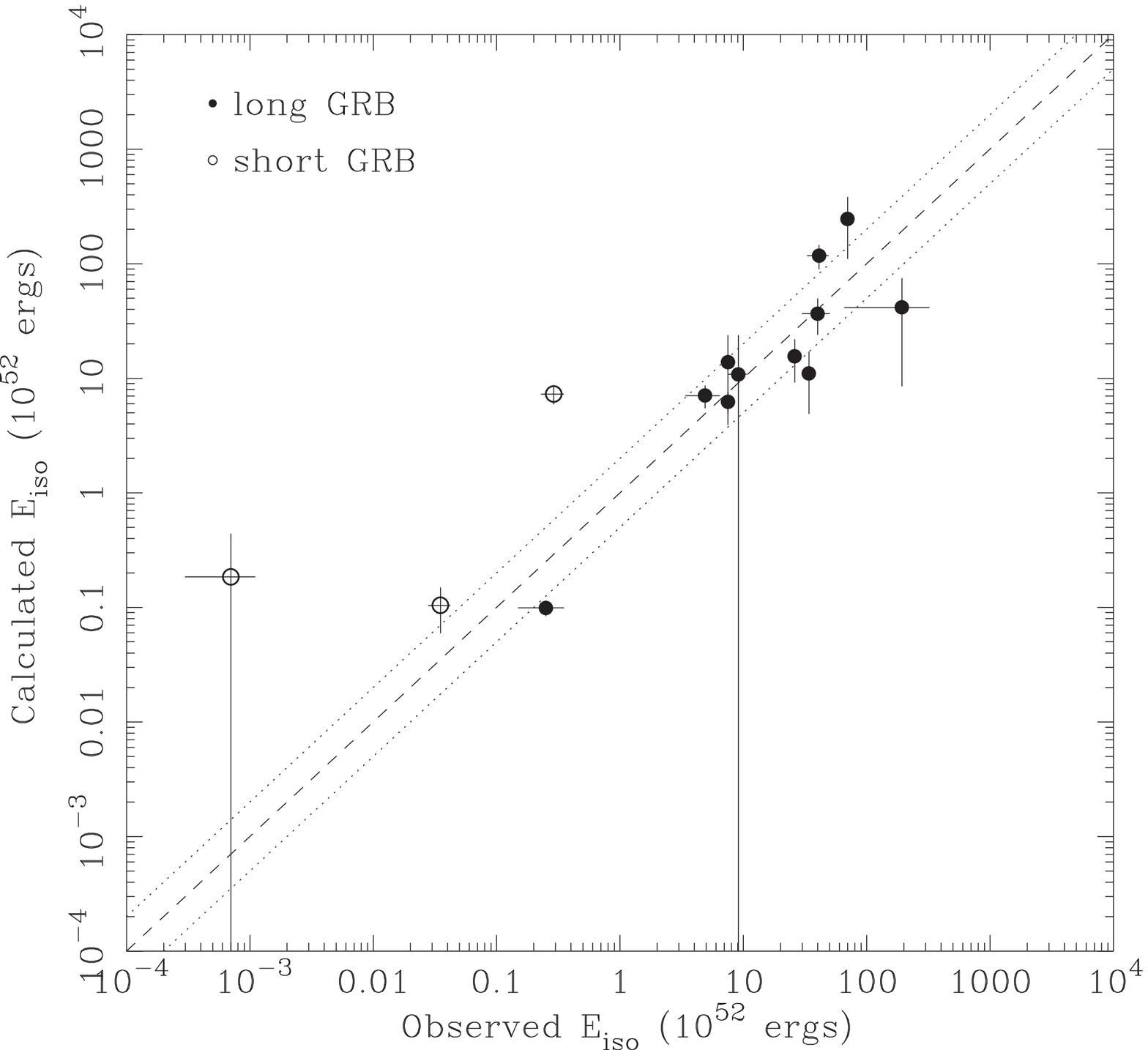}
\caption{Comparison of the isotropic gamma-ray energy of the prompt emission $E_{\gamma, \rm iso}$ 
derived from our calculation method with firm estimated values $E_{\gamma, \rm iso}^{\rm obs}$ (see Section 2.2.2). 
The dashed line is the calculated $E_{\gamma, \rm iso} = E_{\gamma, \rm iso}^{\rm obs}$. 
The dotted lines show difference with a factor of two between between the observed and calculated values. \label{fig:hikaku}}
\end{center}
\end{figure}

From equations (6) and (7), we can obtain the peak energy $E_{\rm p}$.
Then we apply the $E_{\rm p} - E_{\gamma, \rm iso}$ relation (Amati 2006b) to 
estimate $E_{\gamma, \rm iso}$, by assuming $E_{\rm p}$ calculated above. 
For the bursts with observed $E_{\gamma, \rm iso}$ values, we compared $E_{\gamma, \rm iso}$ 
calculated from our method and observed value. Then we confirmed that  
calculated values are approximately consistent with the observed values. 
In Figure \ref{fig:hikaku}, the dotted lines show difference with a factor of two between the observed and calculated values. 
We see that except for short GRB, the observed and calculated values agree within a factor of two. 
We fitted the BAT spectra obtained in Section 2.1 with the single power-law 
($N(E) \propto E^{-\Gamma}$, where $\Gamma$ is the differential photon index).
Then we apply the $E_{\rm p}$ calculation method to these BAT photon indices. 
The results are summarized in Table \ref{table:BATspec}.

\section{Results}

\subsection{$L_{X} - T_{\rm brk}$ relation}

\begin{figure}
\begin{center}
\includegraphics[scale=.40]{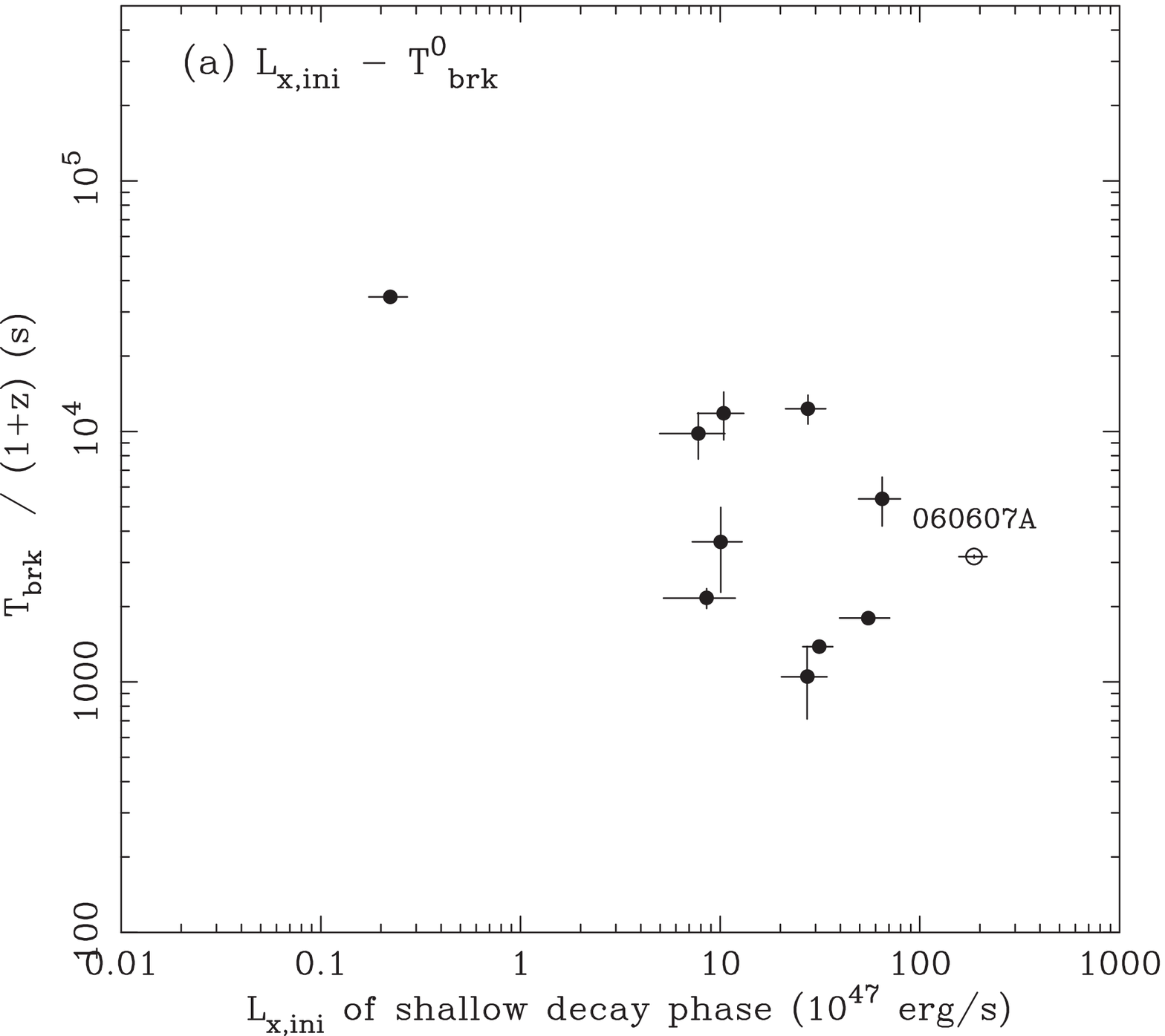}
\includegraphics[scale=.40]{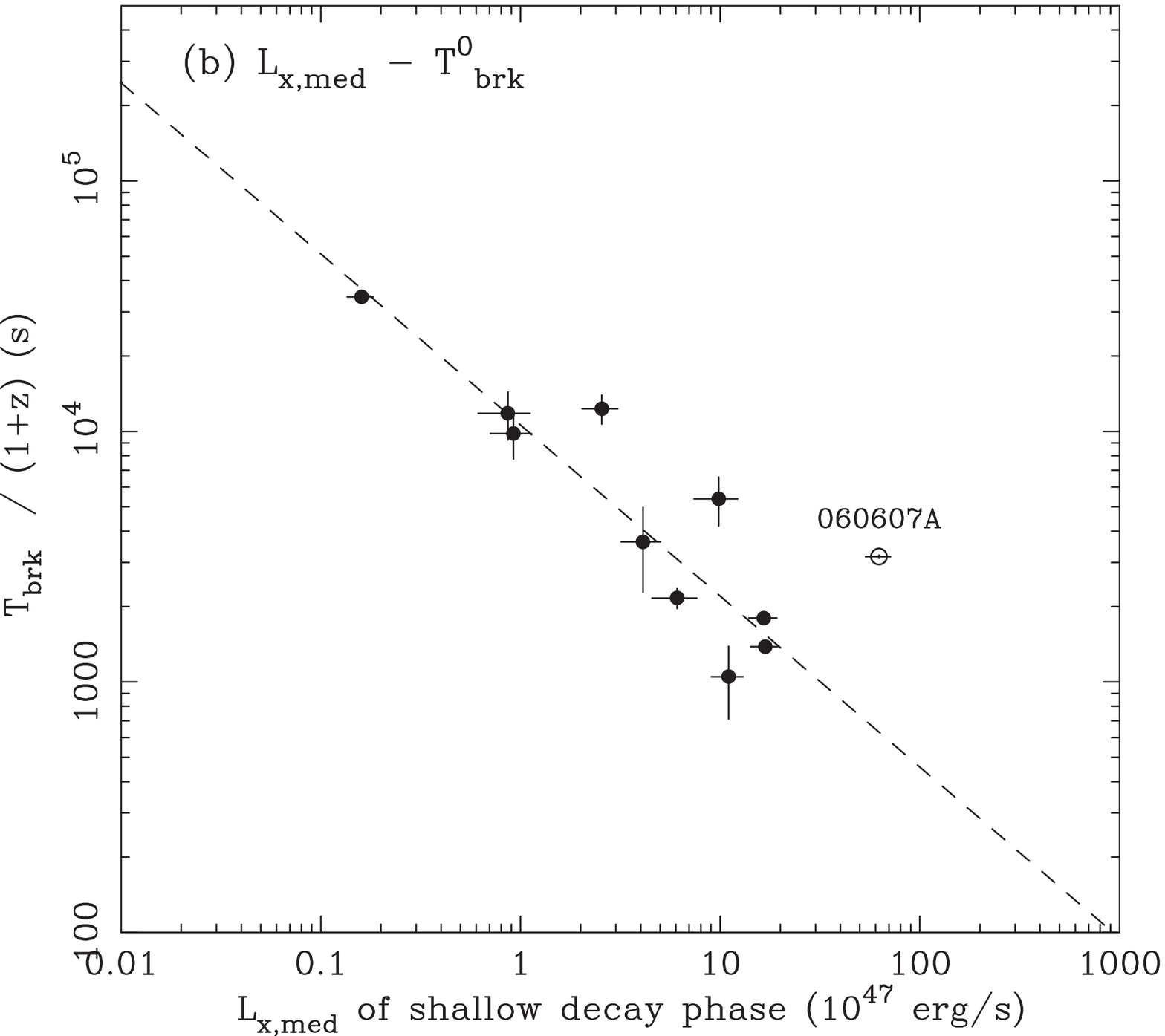}
\includegraphics[scale=.40]{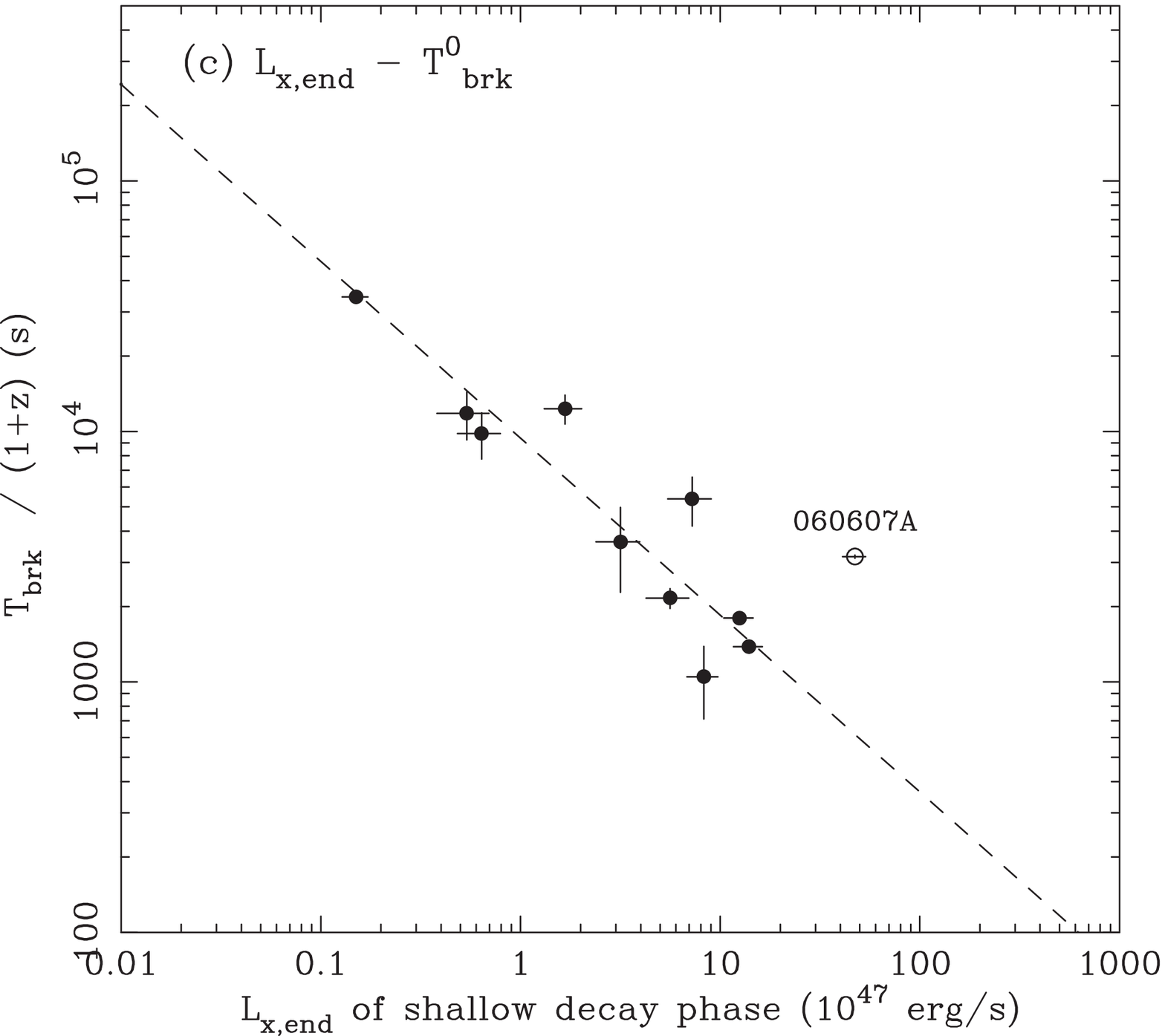}
\caption{Distribution of the intrinsic break time at the shallow-to-normal decay transition in the X-ray light curve
$T^0_{\rm brk} = T_{\rm brk}/(1+z)$ as a function of the isotropic X-ray luminosity of the shallow decay phase 
$L_{X}$ at different epochs: (a) $L_{X, \rm ini}$ at the beginning of the shallow decay, 
(b) $L_{X, \rm med}$ at the median epoch and (c) $L_{X, \rm end}$ at the end. 
The dashed line is the best-fit power-low model for the data (Eq. 8). 
The open circle shows the unusual afterglow of GRB 060607A.\label{fig:Lx-Tbrk}}
\end{center}
\end{figure}

Figure \ref{fig:Lx-Tbrk} shows the distribution of 
the intrinsic break time at the shallow-to-normal decay transition in the X-ray light curve
in the GRB frame $T^0_{\rm brk}$ as a function of the X-ray afterglow luminosities at different epochs: 
$L_{X, \rm ini}$ at the beginning of the shallow decay, 
$L_{X, \rm med}$ at the median epoch and $L_{X, \rm end}$ at the end.
Eleven GRBs in our sample have redshift measurements and XRT data in the start and the end time ($T_{\rm brk}$) 
of the shallow decay phase. 
There are moderately good correlations between afterglow luminosities and $T_{\rm brk}$. 
In these samples  GRB 060607A is unusual since it has an abrupt break at
$T_{\rm brk}$ similar to GRB070110 (Troja et al. 2007) and 
the origin of shallow decay could be different from others (Liang et al. 2007).
Considering GRB060607A as an outlier, 
the correlation coefficient are 0.47, 0.65, 0.65 for 9 degrees of
freedom (10 GRBs) for Figure \ref{fig:Lx-Tbrk} a, b and c, respectively.
The chance probability are 0.17, 0.042, 0.042, respectively. 

When we adopted the power-law model to the $L_{X, \rm med}$ - $T^0_{\rm brk}$ 
and the $L_{X, \rm end}$ - $T^0_{\rm brk}$ relation, the best-fit function are 
\begin{equation}
T^0_{\rm brk}=\frac{T_{\rm brk}}{(1+z)} \hspace*{0.2cm} {\rm (s)} = \left\{
\begin{array}{@{\,}ll}
(10.6\pm0.78) \times 10^3 \biggl(\frac{L_{X}}{10^{47} {\rm ergs \hspace*{0.1cm} s^{-1}}}\biggr)^{-0.68\pm0.03} \hspace*{0.3cm} {\rm for \hspace*{0.3cm} L_{X,\rm med}}, \\
(9.39\pm0.64) \times 10^3 \biggl(\frac{L_{X}}{10^{47} {\rm ergs \hspace*{0.1cm} s^{-1}}}\biggr)^{-0.71\pm0.03} \hspace*{0.3cm} {\rm for \hspace*{0.3cm} L_{X,\rm end}}. \\
\end{array}
\right .\
\end{equation}

\begin{figure}
\begin{center}
\includegraphics[scale=.50]{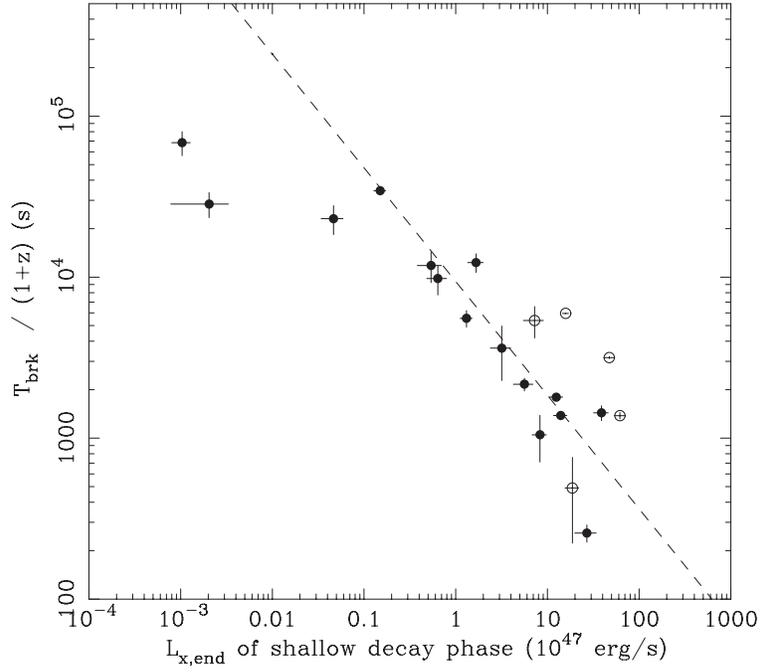}
\caption{Distribution of the intrinsic break time at the shallow-to-normal decay transition in the X-ray light curve
$T^0_{\rm brk} = T_{\rm brk}/(1+z)$ as a function of the isotropic X-ray luminosity 
at the end of the shallow decay $L_{X, \rm end}$ .
The open circles show the unusual afterglows 
which have an abrupt break at $T_{\rm brk}$ or a chromatic X-ray light curve break 
(see Section 3.1). \label{fig:Lend-Tbrk}}
\end{center}
\end{figure}

Figure \ref{fig:Lend-Tbrk} shows $L_{X, \rm end} - T^0_{\rm brk}$ relation for all the bursts 
that have $T_{\rm brk}$ measurements including those missing the observational data 
at the beginning of the shallow decay phase.
We found that two types of the bursts seem to deviate from the $L_{X, \rm end} - T^0_{\rm brk}$ correlation: 
the bursts which have small X-ray luminosity at $T_{\rm brk}$, 
and the bursts which have abrupt or chromatic X-ray light curve breaks 
(GRB 060607A, 050319, 050401 (Panaitescu et al. 2006), 060210 (Stanek et al. 2007), and 060927 (Ruiz-Velasco et al. 2007)). 
These plots are shown as open circles.

\begin{figure}
\begin{center}
\includegraphics[scale=.50]{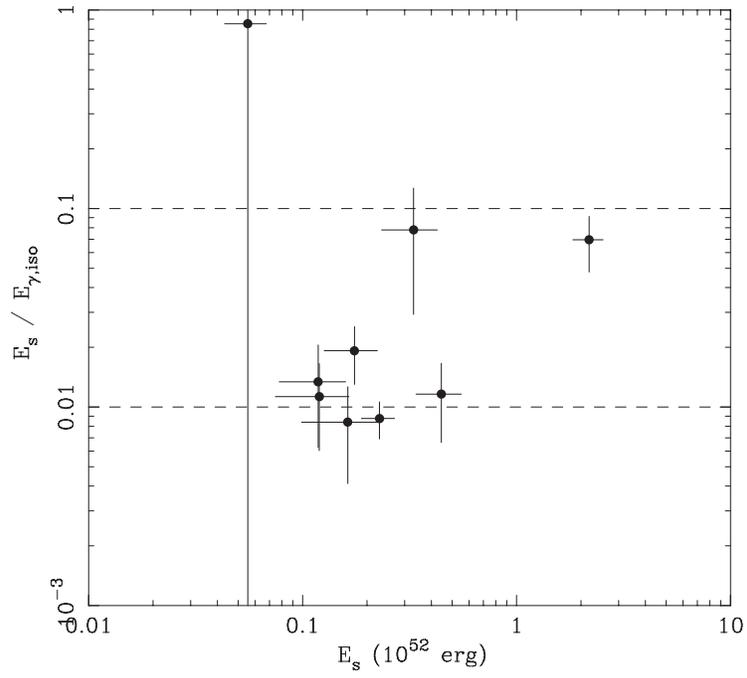}
\caption{Distribution of the integrated energy in the shallow phase ($E_s$) and 
the isotropic gamma-ray energy of the prompt emission $E_{\gamma, \rm iso}$. 
$E_s$ is typically (0.01$\sim$0.1)$E_{\gamma, \rm iso}$. \label{fig:Ex-Tbrk}}
\end{center}
\end{figure}

In Figure \ref{fig:Ex-Tbrk}, we show the integrated energy in the shallow phase ($E_s/E_{\gamma, \rm iso}$) and $E_{\gamma, \rm iso}$. 
We see that $E_s$ is typically (0.01$\sim$0.1) $E_{\gamma, \rm iso}$. This result is similar to Figure 3-d of Liang et al. (2007).

\subsection{$E_{\rm \gamma, iso} - T_{\rm brk}$ relation}

\begin{figure}
\begin{center}
\includegraphics[scale=.50]{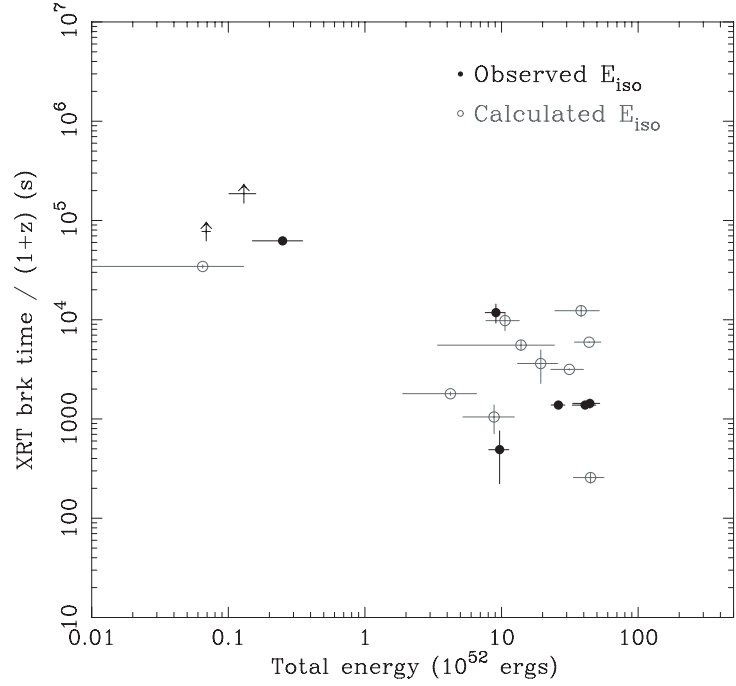}
\caption{Distribution of $T^0_{\rm brk}= T_{\rm brk}/(1+z)$ and the isotropic gamma-ray energy of the prompt emission 
$E_{\gamma, \rm iso}$. 
We found that $T^0_{\rm brk}$ is weakly anti-correlated with $E_{\gamma, \rm iso}$ in logarithmic scale. \label{fig:inhomo}} 
\end{center}
\end{figure}

Figure \ref{fig:inhomo} shows the distribution of $T^0_{\rm brk}$ and 
the isotropic gamma-ray energy of the prompt emission $E_{\gamma, \rm iso}$.
We found that $T^0_{\rm brk}$ is weakly anti-correlated with $E_{\gamma, \rm iso}$ in logarithmic scale
\footnote{In order to check validity of that assumption of $\Gamma_2 = 2.2$, 
we tested the case of $\Gamma_2=2.1$ or 2.5. Although the slope of the plot changed with the $\Gamma_2$ value, 
it did not affect the correlation between $E_{\gamma,\rm iso}$ and $T^0_{\rm brk}$. 
We also plotted only the data with firmly identified $E_{\gamma,\rm iso}$. 
Though the dispersion of the data point is large, there seem to be a trend that the larger $E_{\gamma,\rm iso}$, 
the earlier $T^0_{\rm brk}$.}.
The correlation coefficient is 0.49 for 16 degrees of freedom 
(15 GRBs except GRB 050824 which has only the lower limit for $T^0_{\rm brk}$) 
and the chance probability show value of 0.057. 
This correlation suggests that the larger the isotropic equivalent energy 
the earlier the end time of the shallow decay phase.

Note that Nava et al. (2007) suggested that $T_{\rm a}$ weakly correlates with $E_{\gamma,\rm iso}$ in their sample. 
However, this correlation disappears when considering all bursts of known redshift and $T_{\rm a}$.
$T_{\rm a}$ is obtained by fitting the X-ray light curves with the prompt emission and the afterglow 
component functions (Willingale et al. 2007). 
In their methods, even if the shallow decay phase is not clearly observed in the X-ray light curve, 
they can obtain the $T_{\rm a}$ value. 
On the other hand, we defined the shallow decay phase when the light curve decay index 
is flatter than the canonical value of $\alpha \sim 1$. 
As for the GRBs satisfying this condition, our plot and the plot showed in Nava et al. (2007) 
are approximately consistent. 
Furthermore, Liang et al. (2007) showed that there is no significant correlations 
between $T^0_{\rm brk}$ and $E_{\gamma,\rm iso}$ in their sample. 
They defined the shallow decay phase has $\alpha < 0.7$ since the decay slope of the normal decay 
phase  predicted by the external GRB models is generally greater than 0.7. 
Therefore the GRBs used for the $T^0_{\rm brk} - E_{\gamma,\rm iso}$ correlation test 
are slightly different from ours.   

\section{Estimation of the redshift using  $L_{X, \rm end} - T^0_{\rm brk}$ correlation}

In the previous section we found a moderately good correlation between $T^0_{\rm brk}$ 
and the isotropic X-ray luminosity in the end of the shallow decay phase ($L_{X, \rm end}$) 
as well as the weak correlation between $T^0_{\rm brk}$ and  $E_{\gamma, \rm iso}$. 
The correlation coefficients are not so good to insist the relations. 
Nevertheless, in this section assuming that  $L_{X, \rm end} - T^0_{\rm brk}$ correlation is correct 
we apply the relation to determine the redshifts of GRBs observed by $Swift$ X-Ray Telescope (XRT) 
during the period of January 2005 to December 2006. This is challenging and important since
in $\sim$ 200 GRBs observed by $Swift$ only $\sim$ 50 have the spectroscopically measured redshifts 
so that another distance indicator using only BAT and XRT data is of great value.

We first rewrite equation (8) using only the observed quantities and the unknown redshift $z$ as
\begin{equation}
(1+z)^{0.71\times\Gamma - 1} r(z)^{1.41}
= 4.78\times10^{-4} \times T_{\rm brk}^{-1} \times \biggl( A\times\frac{(E_{\rm u}^0)^{2-\Gamma} - (E_{\rm d}^0)^{2-\Gamma}}{(2-\Gamma)} \biggr)^{-0.71},
\end{equation}

where

\begin{equation}
r(z)=\int_0^z\frac{dz}{\sqrt{\Omega_m(1+z)^3+\Omega_\Lambda}}. 
\end{equation}

Since the left hand side of equation (10) is a monotonically increasing function from zero, there is only one solution of
$z$ for any observed values of $T_{\rm brk}$ and $F_{X, \rm end}$.
We call the redshift obtained by this method as the pseudo $z$
from $L_{X, \rm end} - T^0_{\rm brk}$ correlation.

\begin{figure}
\begin{center}
\includegraphics[scale=.50]{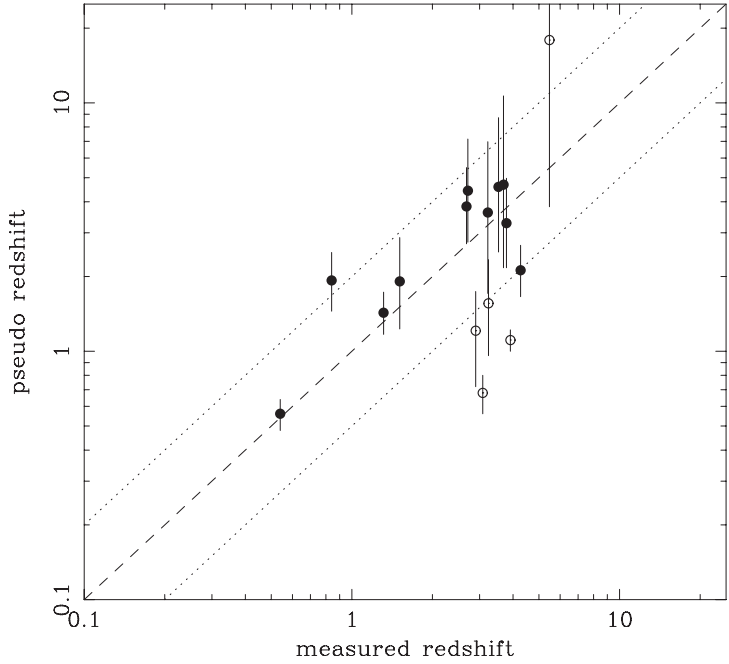}
\caption{Comparison of pseudo redshifts and the observed values except for the burst which have small X-ray luminosity 
at $T_{\rm brk}$.   
The pseudo redshifts were estimated from the correlation between 
the isotropic X-ray luminosity at the end of the shallow decay $L_{X, \rm end}$ and 
the X-ray break times at the shallow-to-normal decay transition in the GRB frame $T^0_{\rm brk}$.
The open circles show the unusual afterglows, which have abrupt/chromatic X-ray light curve breaks. 
The dashed line is the pseudo redshift = measured redshift. 
The dotted lines show factor of two difference between the observed and the pseudo values. \label{fig:z-estimate}}
\end{center}
\end{figure}

In Table \ref{table:z-estimate} we show the list of the pseudo redshifts obtained 
for GRBs observed by $Swift$ X-Ray Telescope (XRT) during the period of January 2005 to December 2006. 
Figure \ref{fig:z-estimate} compares the pseudo redshifts with spectroscopically determined redshifts.  
Although the error bars of the pseudo redshifts are rather large, 
we see that the pseudo redshift determined by our method is a relatively good measure of the redshift of GRB 
for which no spectroscopic information is available. 
The correlation coefficient is 0.58 for 15 degrees of freedom (16 GRBs). 

In Figure \ref{fig:z-estimate}, the dotted lines show difference by a factor of two between the observed and pseudo redshifts.
We see that except for two GRBs, which have abrupt/chromatic X-ray light curve breaks (GRB 060210 and 060607A: different by a factor of three), 
the observed and pseudo redshifts agree within a factor of two so that we may say that our pseudo redshift is the measure of the redshift 
if we allow a factor of two error. 

\begin{figure}
\begin{center}
\includegraphics[scale=.50]{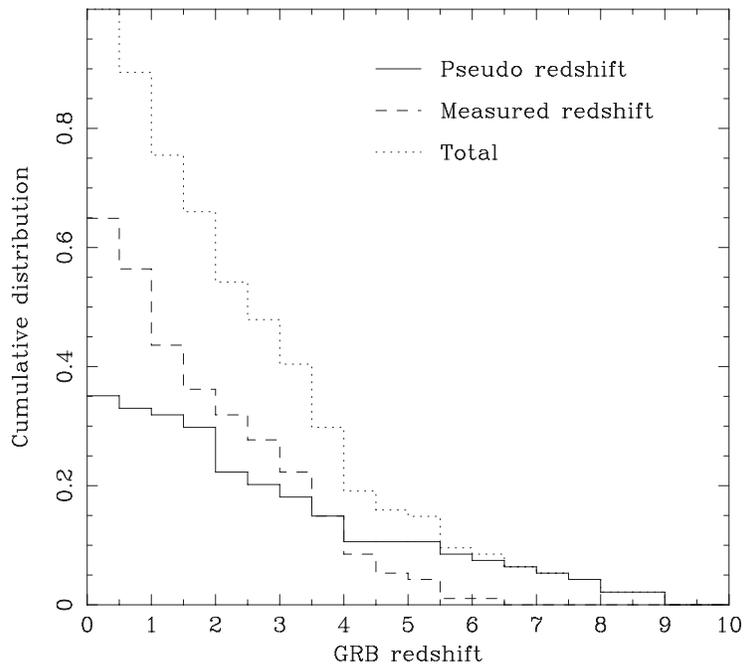}
\caption{Cumulative distribution of the observed (the dashed line) and the pseudo (the solid line) redshifts. 
The dotted line show the total distribution. The mean redshifts are 2.6. \label{fig:z-hist}}
\end{center}
\end{figure}

In Figure \ref{fig:z-hist} we show the cumulative distribution of the spectroscopically measured redshifts ((1), the dashed line), 
the pseudo redshifts for GRBs with no spectroscopically measured redshifts ((2), the solid line) 
and the total ((1)+(2), the dotted line). 
Note that the normalization of the cumulative distribution is the total one for (1) and (2).
The mean redshifts are 2.2 and 2.6 for the observed and the total redshifts, respectively. 
There is a slight difference in the distribution between of the pseudo z and observed z, 
but it is probably due to the difficulty of obtaining spectra at high redshifts. 
Figure \ref{fig:z-hist} suggests that $\sim 15\%$ of GRBs have
redshifts greater than 5. This is consistent 
with the constraints from the optically observed GRBs
(Tanvir \& Jakobsson 2007)  
 
\section{Discussions}

Since $T^0_{\rm brk} - E_{\gamma, \rm iso}$ correlation is weak we here argue mainly the physical implications 
of $T^0_{\rm brk} - L_{X, \rm end}$ correlation as described by $T^0_{\rm brk}\propto L_{X, \rm end}^{-0.7}$. 
Here we note it is reasonable that $T^0_{\rm brk} - L_{X, \rm end}$ correlation 
is better than $T^0_{\rm brk} - L_{X, \rm ini}$ correlation since in the shallow decay phase more energy is emitted 
in the end phase than the initial phase.
Interestingly this correlation is similar to the burning time $t_H$ and the luminosity $L$ relation of the hydrogen main
sequence star ($t_H\propto L^{-0.7}$).  From the theory of stellar evolution  $L$ and $t_H$ mainly 
depend on the mass of the star $M$ as
\begin{eqnarray}
  L\propto M^3,\\
t_H\propto \frac{M}{L}\propto M^{-2}.
\end{eqnarray}
Eliminating $M$ from above two equations we have $t_H\propto L^{-0.67}$.

\subsection{Energy Injection Model}

The energy injection model (Nousek et al. 2006; Zhang et al. 2006; Granot \& Kumar 2006)
is thought to be that energy is injected continuously into the external shock 
so that the flux decay becomes slower than the usual $\propto t^{-1}$. 
The injection may be caused by (1) a long-lived central engine or (2) a short-lived central 
engine ejecting shells with some range of Lorentz factors.

First we consider (1), the long-lived central engine model. 
This scenario requires the central engine to remain active until the end of the shallow decay phase ($T^0_{\rm brk}$), 
which is in many cases $1,000-10,000$ s. 
 Since the X-ray afterglow is a good indicator of the kinetic energy,
the central engine injects energy with the kinetic luminosity
proportional to the X-ray luminosity $L_{\rm kin} \propto L_{\rm X} \propto t^{-\alpha_1}$.
The kinetic energy ($E_{\rm kin}$) of the afterglow increases
as a function of time $\propto t^{1-\alpha_1}$, and 
the observed correlation suggests that the total kinetic energy is
anti-correlated with the lifetime of the central engine,
\begin{eqnarray}
E_{\rm kin,end} \propto 
L_{\rm X,end} T_{\rm brk}^{0} \propto 
\left(T_{\rm brk}^{0}\right)^{-\frac{1}{0.71}+1}
\sim \left(T_{\rm brk}^{0}\right)^{-0.41}.
\end{eqnarray}
where $E_{\rm kin, end}$ is the kinetic energy of afterglow in the end of the shallow decay phase.

Next we consider (2) a short-lived central engine with some range of Lorentz factors of ejected shells. 
After the internal shocks, shells are rearranged so that outer shells are faster 
and inner shells are slower. This configuration may also occur 
if the central engine ejects faster shell earlier. 
Outer shells are slowed down by making the external shock. 
Once the Lorentz factor of the leading shocked shell drops below that of a following slower shell, 
the slower shell catches up with the shocked shell, injecting energy into the forward shock. 
 Since the Lorentz factor of the afterglow is proportional to 
$\Gamma \propto E^{1/8} T^{-3/8} n^{-1/8}$,
the observed correlation suggests that 
the Lorentz factor of the slowest shell, which has almost all energy, is 
nearly proportional to the total kinetic energy,
\begin{eqnarray}
\Gamma_{\rm slow} \propto E_{\rm kin,end}^{1.0} n^{-1/8},
\label{eq:ga}
\end{eqnarray}
where $n$ is the ambient density.

\subsection{Inhomogeneous Jet Model}

In the inhomogeneous jet model, it is assumed that we observe more energetic
components in the GRB jet at later times as the external shock decelerates and
the visible region increases.
The shallow decay phase is produced by the superposition of the afterglow
emission from the off-axis components.
This phase ends when the whole jet is observed in the ring-shaped jet model
(Eichler \& Granot 2006) or when the mini-jets merge and the inhomogeneities
are averaged out in the multiple mini-jets model (Toma et al. 2006).

In the ring-shaped jet model, the X-ray luminosity at the end of the shallow phase
is given by $L_{\rm X, end} \propto E_{\rm kin, iso} (T^0_{\rm brk})^{-1}$, and
the opening angle of the whole jet is determined by 
$\theta_j \propto E_{\rm kin, iso}^{-1/8} (T^0_{\rm brk})^{3/8} n^{1/8}$,
where $E_{\rm kin, iso}$ is the isotropic kinetic energy of the whole jet.
Thus the observed correlation suggests
$\theta_j \propto E_{\rm kin, iso}^{-1.0} n^{1/8}$.
Since the collimation-corrected kinetic energy is calculated as
$E_{\rm kin} \propto E_{\rm kin,iso} \theta_j^2$, we obtain
$E_{\rm kin} \propto \theta_j  n^{1/8}$.

\subsection{Time-dependent Microphysics Model}

This model considers that the microphysical parameters, such as the energy 
fraction that is shared to electrons $\epsilon_{\rm e}$ and magnetic field $\epsilon_{\rm B}$, 
depend on time (Ioka et al. 2006; Fan \& Piran 2006; 
Panaitescu et al. 2006b). 
We usually assume that the micro-physical parameters do not vary and in fact, 
constant $\epsilon_{\rm e}$ and $\epsilon_{\rm B}$ are consistent with the observation 
of late time afterglows (Yost et al. 2003). 
However, the behavior of these parameters in the early time afterglow is not yet known. 

In the model, the microphysical parameters vary in the early afterglow. 
After reaching the equipartition value, the microphysical parameters remain constant 
as observed in the late time afterglow. 
The X-ray luminosity $L_{\rm X}$ is given by the bolometric kinetic luminosity $L$ as 
$L_{\rm X}\sim\epsilon_{\rm e} L$. Since $L\propto E_{\rm kin}t^{-1}$, the shallow X-ray light curve 
$L_{\rm X} \propto t^{-1/2}$ suggests that $\epsilon_{\rm e}$ evolves as $\epsilon_{\rm e} \propto t^{1/2}$ 
(Ioka et al. 2005). 
 The observed correlation suggests that 
the saturation occurs at the Lorentz factor 
$\Gamma_{\rm sat} \propto E_{\rm kin,end}^{1.0} n^{-1/8}$
as in equation (\ref{eq:ga}).

\subsection{Pulsar Model}

Troja et al. (2007) showed that the abrupt drop of the X-ray light curve 
observed in GRB 070110 cannot be explained by an external shock as the origin of the shallow 
decay phase and implies the long-lived central engine.
Furthermore, they suggest that the shallow decay phase might be powered by a spinning down central engine, 
possibly a millisecond pulsar.
Motivated by this suggestion let us consider that the slowly changing usual shallow
decay is also due to the pulsar activity. 
If the dipole magnetic field is constant, the luminosity of the pulsar decreases as $\propto t^{-2}$,
so that we need the increase of the dipole magnetic field to interpret the shallow decay phase. 
 From the total energy conservation, we have
\begin{eqnarray}
\frac{d}{dt}\left(\frac{1}{2}I\Omega^2+\frac{B^2R^3}{6}\right)=-L(t),
\end{eqnarray}
where we assume that the rotational energy goes into the magnetic energy
by some unknown mechanism satisfying the observed luminosity,
\begin{eqnarray}
L(t)=\frac{B^2\Omega^4R^6}{6c^3}
=L_0 \left(\frac{t}{t_0}\right)^{-\alpha_1},
\label{eq:dipole}
\end{eqnarray}
Integrating the above equations from the beginning of the shallow decay phase $t_0$ to $T^0_{\rm brk}$,
we have
\begin{equation}
\frac{1}{2}I(\Omega^2-\Omega_0^2)+\frac{(B^2-B^2_0)R^3}{6}
=-L_{\rm X,end} T^0_{\rm brk} \frac{1}{1-\alpha_1} 
\left[1-\left( \frac{t_0}{T^0_{\rm brk}}\right)^{1-\alpha_1}  \right]
\end{equation}
In this model $T^0_{\rm brk}$ is essentially determined by the
total energy conservation as
\begin{equation}
 \frac{1}{2}I\Omega_0^2\sim L_{X, \rm end}T^0_{\rm brk}/(1-\alpha_1)=E_s.
\end{equation}
Therefore the energy of the shallow decay phase 
$E_s$ is essentially the total energy of the initial
pulsar. Figure \ref{fig:Ex-Tbrk} suggests that the initial rotational period is
1ms$\sim$ 3ms, which is an appropriate value in this model.
Using equation (\ref{eq:dipole}) and 
$T^0_{\rm brk}$-$L_{X, \rm end}$ relation we have
\begin{equation}
 B_0^2\Omega_0^4\propto 
\left(\frac{T^0_{\rm brk}}{t_0}\right)^{\alpha_1}(T^0_{\rm brk})^{-\frac{1}{0.71}}.
\end{equation}
Since $\Omega_0$ is known from $E_s$, the above equation tells
us that the initial strength of the magnetic field determines
$T^0_{\rm brk}$.

\section{Summary}

From our observational results, we found that  the intrinsic break time 
at the shallow-to-normal decay transition 
in the X-ray light curve $T^0_{\rm brk}$ is moderately well correlated 
with the isotropic X-ray luminosity
in the end of the shallow decay phase ($L_{X, \rm end}$) as 
$T^0_{\rm brk}=(9.39\pm0.64) \times 10^3 {\rm s}({L_{X,\rm end}}/{10^{47} {\rm ergs\cdot s^{-1}}})^{-0.71\pm0.03}$, 
while $T^0_{\rm brk}$ is weakly correlated with the isotropic gamma-ray energy of the prompt emission $E_{\gamma, \rm iso}$.
Using this relation we have determined the pseudo redshifts of 33 GRBs and found that the distribution 
of the pseudo redshifts is similar to that of spectroscopically determined redshifts.
Since the $T^0_{\rm brk}-L_{X, \rm end}$ relation does not have an intrinsic redshift degeneracy, 
we can determine the redshift of the GRB uniquely. 
 The $T^0_{\rm brk}-L_{X, \rm end}$ relation does not 
require the parameters of the prompt emission so that 
it may be useful to determine the redshift of $Swift$ GRBs
since the energy band of $Swift$ is typically below 
the peak energy of the prompt emission.
Our results suggest that $\sim 15\%$ of GRBs have $z > 5$. 
This means an exciting possibility  such that
 the redshift is estimated in advance of deep follow-ups
and possible high redshift GRBs ($z > 6.3$) might be selected 
for detailed observations and identified finally 
in near future .

We discussed the implications of the $T^0_{\rm brk} - L_{X, \rm end}$ relation for some theoretical models
recently proposed to explain the shallow decay light curve.
In each model, we obtain an additional condition for the models to be satisfied from 
the $T^0_{\rm brk} - L_{X, \rm end}$ relation.
Other models including two-component jet (Granot \& Kumar 2006; Jin et al. 2007),
dust scattering (Shao \& Dai 2007), and relativistic wind bubbles produced by the interaction
of an ultra-relativistic electron-positron-pair wind with an outwardly expanding fireball
(Dai 2004; Yu \& Dai 2007), have also been proposed, but the detailed discussion for these models
are beyond the scope of this paper.

\acknowledgments

We are grateful to T. Sakamoto, G. Sato, and all the members of the $Swift$ Team 
for their technical guidance of the analysis of $Swift$ BAT and XRT data.
R.S. is supported by the Research Fellowships for Young Scientists (2005-2007) 
of the Japan Society for the Promotion of Science.
This work is supported in part by the Grant-in-Aid from the 
Ministry of Education, Culture, Sports, Science and Technology
(MEXT) of Japan, No.18740147 (K.I.) and No.19540283,No.19047004,No. 19035006(T.N.).


%% in parentheses, after the keyword, but they will not be verified.



% Figure %

\clearpage

% Table %

\begin{table}
\begin{center}
\
\caption{Temporal Parameters of the shallow-to-normal decay phase in the X-ray light curves for known-redshift GRBs. 
Parameters of $t_{\rm sta}$ and $t_{\rm sto}$ show the fitting range and the fitting model is a broken power-law (Eq. 1). 
\label{table:LCfit}}
\begin{tabular}{lccccccc}
\tableline\tableline
GRB     & z & $t_{\rm sta}$ & $t_{\rm sto}$  & $\alpha_1$ & $\log T_{\rm brk}$ & $\alpha_2$ & Reduced $\chi^2$ \\
        &   &     (s)       &      (s)       &            &        (s)         &            &   (d.o.f)        \\   
\tableline
050319  & 3.24   &  389 & 417809 & 0.49$\pm$0.03 & 4.36$\pm$0.10 & 1.27$\pm$0.26 & 1.02 (21) \\
050401  & 2.90   & 1020 &  20108 & 0.43$\pm$0.08 & 3.73$\pm$2.39 & 1.42$\pm$0.55 & 1.01 (12) \\
050505  & 4.27   & 2832 &  44835 & 0.15$\pm$0.15 & 3.88$\pm$2.90 & 1.17$\pm$0.05 & 0.62 (17) \\
050814  & 5.3    & 5759 & 215431 & 0.63$\pm$0.04 & 4.89$\pm$0.06 & 1.99$\pm$0.82 & 1.19 (15) \\
050824  & 0.83   & 6091 & 603407 & 0.52$\pm$0.04 & - & - & 1.99 (9) \\
051016B & 0.9364 & 4055 & 278391 & 0.73$\pm$0.06 & 4.65$\pm$0.10 & 1.36$\pm$0.16 & 0.68 (22) \\
060115  & 3.53   &  749 & 209997 & 0.70$\pm$0.04 & 4.73$\pm$0.10 & 1.80$\pm$0.53 & 1.10 (24) \\
060202  & 0.783  & 4942 & 745686 & 0.87$\pm$0.02 & - & - & 1.07 (46)\\
060210  & 3.91   & 3849 & 896348 & 0.92$\pm$0.07 & 4.47$\pm$1.82 & 1.37$\pm$0.03 & 0.97 (38) \\
060502  & 1.51   &  281 & 676831 & 0.54$\pm$0.05 & 4.39$\pm$3.71 & 1.19$\pm$0.05 & 0.99 (17) \\
060526  & 3.221  &  836 & 220231 & 0.41$\pm$0.06 & 4.18$\pm$3.75 & 1.46$\pm$0.24 & 0.77 (13) \\
060604  & 2.68   & 3470 & 208383 & 0.35$\pm$0.08 & 4.31$\pm$3.38 & 1.28$\pm$0.09 & 0.99 (18) \\
060605  & 3.78   &  230 &  74165 & 0.42$\pm$0.04 & 3.93$\pm$2.54 & 2.01$\pm$0.07 & 1.14 (25) \\
060607  & 3.08   &  493 &  48519 & 0.42$\pm$0.02 & 4.11$\pm$2.28 & 3.62$\pm$0.07 & 2.12 (36) \\
060614  & 0.125  & 4430 & 555793 & 0.11$\pm$0.04 & 4.85$\pm$3.70 & 2.24$\pm$0.13 & 2.20 (36) \\
060714  & 2.71   &  283 & 284999 & 0.47$\pm$0.15 & 3.59$\pm$3.10 & 1.20$\pm$0.04 & 0.98 (27) \\
060729  & 0.54   &  681 & 180616 & 0.09$\pm$0.01 & 4.72$\pm$3.28 & 1.18$\pm$0.05 & - \\
060906  & 3.686  &  404 & 139753 & 0.13$\pm$0.08 & 4.01$\pm$0.04 & 1.76$\pm$0.13 & 0.74 (12) \\
060908  & 2.43   &   80 &  92436 & 0.67$\pm$0.05 & 2.81$\pm$0.01 & 1.41$\pm$0.04 & 1.08 (27) \\
060927  & 5.6    &   70 & 191186 & 0.63$\pm$0.20 & 3.51$\pm$3.25 & 1.69$\pm$0.31 & 0.57 (10) \\
061121  & 1.314  &  192 &  92548 & 0.30$\pm$0.23 & 3.51$\pm$2.27 & 1.25$\pm$0.02 & 1.35 (67) \\
\tableline
\end{tabular}
\end{center}
\end{table}

\begin{table}
\begin{center}
\caption{GRB samples used to study the X-ray luminosity and energy in the shallow decay phase. 
$L_{X}$ is given in units of $10^{47}$ erg s$^{-1}$, 
and $E_{s}$ is given in units of $10^{51}$ erg s$^{-1}$. \label{table:Lx}}
\begin{tabular}{lcccccc}
\tableline\tableline
GRB     & $\Gamma$ & $L_{X, \rm ini}$  & $L_{X, \rm med}$  & $L_{X, \rm end}$ & $E_{\rm s}$ \\
        &          & ($10^{47}$ erg/s) & ($10^{47}$ erg/s) & ($10^{47}$ erg/s)& ($10^{51}$ erg/s) \\ 
        &          &                   &                   &                  &         \\      
\tableline
050319  & 2.21$\pm$0.15 & 64.7$\pm$15.3     & 9.82$\pm$2.41     & 7.24$\pm$1.77      & 6.69$\pm$1.79 \\
050814  & 2.17$\pm$0.15 & 27.5$\pm$6.3      & 2.55$\pm$0.52     & 1.67$\pm$0.32      & 4.45$\pm$1.06 \\
060115  & 1.84$\pm$0.09 & 10.4$\pm$2.7      & 0.865$\pm$0.251   & 0.539$\pm$0.156    & 1.75$\pm$0.49 \\
060502  & 2.07$\pm$0.13 & 7.77$\pm$2.79     & 0.922$\pm$0.215   & 0.639$\pm$0.155    & 1.20$\pm$0.45 \\
060526  & 2.35$\pm$0.17 & 10.1$\pm$2.8      & 4.11$\pm$0.91     & 3.17$\pm$0.78      & 1.63$\pm$0.64 \\
060605  & 2.29$\pm$0.13 & 55.1$\pm$15.5     & 16.6$\pm$2.7      & 12.5$\pm$2.1       & 3.30$\pm$0.96 \\
060607A & 1.80$\pm$0.09 & 186.$\pm$30.      & 62.4$\pm$8.9      & 47.3$\pm$6.2       & 21.8$\pm$3.5  \\
060714  & 2.91$\pm$0.17 & 27.3$\pm$7.0      & 11.0$\pm$2.0      & 8.28$\pm$1.46      & 1.18$\pm$0.41 \\
060729  & 2.16$\pm$0.07 & 0.223$\pm$0.049   & 0.160$\pm$0.024   & 0.150$\pm$0.022    & 0.554$\pm$0.122\\
060906  & 2.29$\pm$0.22 & 8.55$\pm$3.35     & 6.10$\pm$1.54     & 5.61$\pm$1.35      & 1.31$\pm$0.52 \\
061121  & 2.21$\pm$0.08 & 31.3$\pm$5.3      & 16.8$\pm$2.6      & 14.0$\pm$2.3       & 2.29$\pm$0.40 \\
\tableline
050401  & 2.10$\pm$0.06 &     -           &         -         & 61.8$\pm$8.4       & - \\
050505  & 2.09$\pm$0.06 &     -           &         -         & 38.9$\pm$7.0       & - \\
051016B & 2.01$\pm$0.12 &     -           &         -         & 0.047$\pm$0.012    & - \\
060210  & 2.25$\pm$0.05 &     -           &         -         & 15.8$\pm$1.2       & - \\
060604  & 1.97$\pm$0.08 &     -           &         -         & 1.31$\pm$0.19      & - \\
060614  & 2.01$\pm$0.09 &     -           &         -         & 0.0010$\pm$0.0002  & - \\
060908  & 2.37$\pm$0.19 &     -           &         -         & 26.9$\pm$7.1       & - \\
060927  & 2.09$\pm$0.16 &     -           &         -         & 18.7$\pm$3.2       & - \\
\tableline
\end{tabular}
\end{center}
\end{table}

\begin{table}
\begin{center}
\caption{Spectral characteristics of the $Swift$ BAT GRBs used to study the $E_{\gamma,\rm iso} - T_{\rm brk}$ correlation. 
Fitting model is a single power-law $N(E)\propto E^{-\Gamma}$. The reference of redshift can be found at 
$http://heasarc.gsfc.nasa.gov/docs/swift/archive/grb_table/$.\label{table:BATspec}} 
\begin{tabular}{lcccccc}
\tableline\tableline
GRB      & $\Gamma$      & BAT mean flux             & Reduced $\chi^2$ & $(1+z)E_{\rm p}$ & $E_{\gamma, \rm iso}$  & $z$ \\
         &               & ($15-150$ keV)            &  (d.o.f)         &  (keV)           & ($10^{52}$ erg)        &     \\
         &               & $10^{-8}$ erg cm$^{-2}$ s$^{-1}$ &           &                  &                        &     \\
\tableline
050319   & 2.12$\pm$0.12 & 9.18$_{-0.62}^{+0.63}$    & 0.44 (21) & - & -                              & 3.24 \\
050814   & 1.86$\pm$0.11 & 2.24$_{-0.16}^{+0.15}$    & 0.86 (21) & 712$\pm$146      & 38.3$\pm$13.7   & 5.3 \\
051016B  & 2.55$\pm$0.22 & 1.42$_{-0.16}^{+0.17}$    & 1.13 (16) & - & -                              & 0.9364 \\
060202   & 2.00$\pm$0.15 & 1.04$_{-0.34}^{+0.53}$    & 1.51 (12) & - & -                              & 0.783 \\
060210   & 1.67$\pm$0.05 & 6.45$\pm$0.20             & 0.73 (21) & 770$\pm$94       & 43.8$\pm$9.6    & 3.91 \\
060502   & 1.32$\pm$0.04 & 11.6$\pm$0.3              & 1.19 (56) & 339$\pm$53       & 10.6$\pm$2.9    & 1.51 \\
060526   & 1.76$\pm$0.10 & 3.99$\pm$0.26             & 0.85 (21) & 480$\pm$90       & 19.4$\pm$6.3    & 3.221 \\
060604   & 1.87$\pm$0.24 & 3.23$_{-0.49}^{+0.50}$    & 0.70 (21) & 396$\pm$173      & 13.9$\pm$10.5   & 2.68 \\
060605   & 1.48$\pm$0.13 & 0.818$\pm$0.067           & 1.79 (21) & 199$\pm$63       & 4.23$\pm$2.34   & 3.78 \\
060607   & 1.32$\pm$0.04 & 6.75$\pm$0.15             & 1.10 (21) & 635$\pm$96       & 31.4$\pm$8.4    & 3.082 \\
060714   & 1.62$\pm$0.10 & 2.88$\pm$0.18             & 0.83 (21) & 304$\pm$71       & 8.83$\pm$3.61   & 2.711 \\
060729   & 1.75$\pm$0.52 & 0.609$_{-0.193}^{+0.202}$ & 0.60 (12) & 17.6 $\pm$10.2   & 0.065$\pm$0.065 & 0.54 \\
060906   & 2.26$_{-0.38}^{+0.46}$ & 1.14$_{-0.26}^{+0.27}$ & 1.22 (12) & - & -                        & 3.686 \\
060908   & 1.25$\pm$0.03 & 13.4$\pm$0.3              & 1.09 (27) & 813$\pm$152      & 44.9$\pm$11.3   & 2.43 \\
\tableline
\end{tabular}
\end{center}
\end{table}

\begin{deluxetable}{lccc}
\tabletypesize{\footnotesize}
\tablecaption{Redshifts estimated by $L_{X} - T_{\rm brk}$ relation.
The reference of observed redshift can be found at $http://heasarc.gsfc.nasa.gov/docs/swift/archive/grb_table/$. 
The pseudo redshift ($\ast$) obtained by another method can be found at $http://cosmos.ast.obs$-$mip.fr/projet/catalog_pz.php$. \label{table:z-estimate}}
\tablewidth{0pt}
\tablehead{
\colhead{GRB} & \colhead{Pseudo $z$ (our results)} & \colhead{Observed $z$} & \colhead{Pseudo $z$($\ast$)}
}
\startdata
050128  & $2.52_{-0.69}^{+0.86}$  &        &\\
050319  & $1.56_{-0.60}^{+0.78}$  & 3.24   &\\
050401  & $1.21_{-0.49}^{+0.53}$  & 2.90   &\\
050505  & $2.12_{-0.46}^{+0.55}$  & 4.27   &\\
050607  & $>$ 3.67                & ($z<5$)&\\
050701  & $5.77_{-5.36}^{-}$      &      &\\
050712  & $1.67_{-0.61}^{+1.22}$  &      &\\
050713A & $2.51_{-1.02}^{+1.44}$  & (0.4-2.6)&\\
050713B & $1.64_{-0.50}^{+0.64}$  &      &\\
050802  & $1.97_{-0.37}^{+0.47}$  & ($z<1.2$, 1.71?)&\\
050803  & $1.52_{-0.31}^{+0.36}$  &      &\\
050814  & $2.80_{-1.08}^{+1.72}$  & (5.3$\pm$0.3 (photometric))&\\
050822  & $2.49_{-0.76}^{+1.29}$  &      &\\
050915A & -                       &      &\\
050922B & $1.32_{-0.71}^{+0.90}$  &      &\\
051008  & $1.88_{-0.44}^{+0.54}$  &      & $z<0.36$, 5.2$\pm$2.2\\
051016B & $2.85_{-1.09}^{+1.76}$  & 0.9364 &\\
051109B & $8.76_{-7.50}^{+6.53}$  & (0.08?)&\\
060105  & $0.77_{-0.18}^{+0.19}$  &      & 4.0$\pm$1.3\\
060108  & $5.78_{-4.18}^{-}$      & ($<2.7$)&\\
060109  & $3.58_{-1.69}^{+3.82}$  &      &\\
060111B & $6.87_{-3.41}^{+6.19}$  &      &\\
060115  & $4.59_{-2.08}^{+4.13}$  & 3.53 &\\
060203  & $7.41_{-3.62}^{+8.45}$  &      &\\
060204B & $3.29_{-1.37}^{+2.40}$  & ($z<4$)& 3.1$\pm$1.1\\
060210  & $1.11_{-0.11}^{+0.11}$  & 3.91 &\\
060219  & $5.46_{-5.46}^{+11.53}$ &      &\\
060306  & $3.89_{-2.79}^{+7.39}$  &      &\\
060312  & -                       &      &\\
060319  & $3.32_{-1.81}^{+4.58}$  &      &\\
060323  & $>$ 8.81                &      &\\
060413  & $0.41_{-0.11}^{+0.10}$  &      &\\
060421  & $>$ 0.08                &      &\\
060502  & $1.91_{-0.68}^{+0.96}$  & 1.51 &\\
060507  & $7.87_{-4.40}^{-}$      &      &\\
060526  & $3.62_{-1.91}^{+3.36}$  & 3.221&\\
060604  & $3.83_{-1.12}^{+1.16}$  & 2.68 &\\
060605  & $3.28_{-1.11}^{+1.69}$  & 3.78 &\\
060607  & $0.68_{-0.12}^{+0.12}$  & 3.082&\\
060708  & $5.35_{-2.81}^{+6.92}$  & (z$\sim$1.8, $z<2.3$)&\\
060714  & $4.43_{-1.68}^{+2.72}$  & 2.711&\\
060719  & $3.67_{-1.97}^{+3.28}$  &      &\\
060729  & $0.56_{-0.08}^{+0.08}$  & 0.54 &\\
060805  & $>$ 6.04                &      &\\
060807  & $3.07_{-1.39}^{+2.68}$  &      &\\
060813  & $1.77_{-0.33}^{+0.38}$  &      & 2.38$\pm$0.40\\
060814  & $1.93_{-0.48}^{+0.57}$  & 0.84 &\\
060906  & $4.69_{-2.52}^{+5.98}$  & 3.686&\\
060908  & $7.91_{-4.23}^{+10.26}$ & 2.43 &\\
060927  & $17.9_{-14.1}^{-}$      & 5.47 &\\
061021  & $2.15_{-0.42}^{+0.49}$  & ($z<2.0$)&\\
061121  & $1.43_{-0.26}^{+0.30}$  & 1.314&\\
\enddata
\end{deluxetable}
\clearpage
 
\end{document}